# Regulation of black-hole accretion by a disk wind during a violent outburst of V404 Cygni

T. Muñoz-Darias[1,2,★] J. Casares[1,2,3], D. Mata Sánchez[1,2], R. P. Fender[3], M. Armas Padilla[1,2,4],

M. Linares[1,2,5], G. Ponti[6], P. A. Charles[7,3], K. P. Mooley[3], J. Rodriguez[8]

[1]Instituto de Astrofísica de Canarias, E-38205 La Laguna, S/C de Tenerife, Spain

[2]Departamento de Astrofísica, Universidad de La Laguna, E-38206 La Laguna, S/C de Tenerife, Spain

[3]Department of Physics, Astrophysics, University of Oxford, Denys Wilkinson Building, Keble Road, Oxford OX1 3RH, UK

[4]Department of Astronomy, Kyoto University, Kyoto 606-8502, Japan

[5]Institutt for fysikk, NTNU, Trondheim, Norway

[6]Max-Planck-Institut für extraterrestrische Physik, Giessenbachstrasse 1, D-85748 Garching bei München, Germany

[7]School of Physics and Astronomy, University of Southampton, Southampton, Hampshire SO17 1BJ, UK

[8]Laboratoire AIM, UMR 7158, CEA/CNRS/Université Paris Diderot,CEA DRF/IRFU/SAp, 91191, Gif-sur-Yvette, France

★e-mail address: teo.munoz-darias@iac.es

**Accretion of matter onto black holes is universally associated with strong radiative feedback[1] and powerful outflows[2]. In particular, black hole transients[3] show outflows whose properties[4] are strongly coupled to those of the accretion flow. This includes X-ray winds of ionized material, expelled from the accretion disc encircling the black hole, and collimated radio jets[5,6]. Very recently, a distinct optical variability pattern has been reported in the transient black hole transient V404 Cyg, and interpreted as disrupted mass flow into the inner regions of its large accretion disc[7]. Here, we report on the discovery of a sustained outer accretion disc wind in V404 Cyg, which is unlike any seen previously. We find that the outflowing wind is neutral, has a large covering factor, expands at 1% of the speed of light and triggers a nebular phase once accretion sharply drops and the ejecta become optically thin. The large expelled mass (> $10^{-8}$ M$_\odot$) indicates that the outburst was prematurely ended when a sizeable fraction of the outer disc was depleted by the wind, detaching the inner regions from the rest of the disc. The luminous, but brief, accretion phases shown by transients with large accretion discs[2] imply that this outflow is most likely a new fundamental ingredient regulating mass accretion onto black holes.**

The X-ray binary V404 Cyg (GS 2023+338) is a confirmed stellar-mass BH[8] with a precisely determined distance of 2.4 kpc[9]. Following 25 years of quiescence, the Swift mission detected renewed activity on Jun 15, 2015[10], initiating a 2-week period of intensely violently variable emission across all wavelengths[11,12]. Our high signal-to-noise GTC optical spectra covering the entire X-ray/radio active phase (~15 days) show that, contemporaneously with radio jet emission, continuous ejections of neutral material at ~0.01c are present from low-level accretion phases (<1% of the Eddington luminosity; $L_{EDD}$) to the X-ray peak (Methods; Fig. 1, ED Fig. 1). These are observed in hydrogen (Balmer) and helium (He I) emission lines as deep P-Cyg profiles throughout the outburst[13], and extremely broad wings once the X-ray and radio fluxes decay. P-Cyg profiles result from resonant scattering in an expanding outflow with a spherical geometry or at least sustaining a large solid angle[14, 15] (Methods). Among a dozen transitions showing this feature, the deepest are seen in the He I-5876 emission line, which is used as a reference for this study (see ED Fig. 2.).

The strongest P-Cyg profiles are witnessed during days 1 to 6 (Fig. 1 and Fig. 2 for the evolution of the profiles during day 2; Methods), when the X-ray luminosity is typically $10^3$ times fainter than the ~$L_{EDD}$ flares displayed later in the outburst[7,11] (ED Fig.1). Blue-shifted absorptions are as deep as 30% below the continuum level and we measure terminal velocities in the range $V_T$=1,500 – 3,000 km s$^{-1}$ (Fig. 1, Fig 2, ED Fig 2; ED Fig. 3). Symmetric red-shifted (i.e. positive velocity) outflow emission, completely detached from the accretion disc line component, is sometimes evident (see Fig. 2 from minute 60 onwards). In order to trace the ionization state of the outer disc we computed the line flux ratio $I_{Ratio}$ = He II-4686 / $H_\beta$ and find $I_{Ratio}$ <0.5 when P-Cyg absorptions are deepest (ED Fig.1; Methods). Through the outburst, both the X-ray and optical emission are characterized by the presence of short and long flaring activity[7,11]. During these flaring episodes $I_{Ratio}$ increases while the P-Cygni profiles become weaker, subsequently recovering their pre-flare strength when the X-ray flux and $I_{Ratio}$ drop (Fig. 2). This indicates that the detection of P-Cyg absorptions is driven by ionization effects. Indeed, on days 7 to 10 much shallower absorptions (only ~2 % below the continuum level) are witnessed as the system enters the brightest phase of the outburst and $I_{Ratio}$ always becomes larger than unity (Fig. 1 and ED Fig.1). Furthermore, we note that the Hα profile is very asymmetric during the whole outburst providing further indication of the ubiquitous presence of wind outflows during our observations (ED Fig 4; Methods).

The low temperature ($T$) required to have both neutral hydrogen ($T$ <$10^4$ K) and helium ($T$ < $3 \times 10^4$ K) places the wind-launching radius ($R_l$) at the outer accretion disc regardless of the wind-launching mechanism. On the other hand, the low luminosity associated with the deepest P-Cyg profiles rules out radiation pressure winds driven by Thomson scattering. The thermal wind scenario[16], in which $V_T$ roughly corresponds to the escape velocity at $R_l$, is able to reproduce our observations. Using $V_T$=1,500 – 3,000 km s$^{-1}$, we obtain $R_l$ = 1.5 – 6 × $10^5$ km, which corresponds to disc temperatures in the range ~5,000 – 30,000 K for mass accretion rate within 0.001 – 0.1 $L_{EDD}$, respectively (Methods). A crude estimation of the mass loss rate associated with the most conspicuous profiles (day 2) suggests $\dot{M}_{out}$ > $10^{-13}$ M$_\odot$ yr$^{-1}$ (Methods). This lower limit only accounts for neutral matter outflows, whilst a wind of ionized material could be also launched. A hot wind with $V_T$ up to ~ 4,000 km s$^{-1}$ is indeed detected by the only Chandra pointing performed during the outburst[17].

The second signature of the high-velocity wind is a short *nebular phase* witnessed at the end of the outburst. Following a sharp drop by a factor of ~$10^3$ in the X-ray, optical and radio luminosity from the major flares ending the brightest phase of the outburst on days 9—10, the Balmer lines became unprecedentedly intense for a black hole, showing equivalent widths up to ~ 2,000 Å (Hα; ED Fig. 1). They sit on extended wings reaching similar velocities to the $V_T$ observed in the P-Cyg profiles (± 3,000 km s$^{-1}$; inset in Fig. 3). A forest of broad emission lines, such as Si II and Fe II also become apparent (Fig. 3) while the Balmer decrement (*BD;* Methods) increases up to ~6, as compared to ~2.5 observed earlier (ED Fig. 1, 5). High values of *BD* are associated with nebulosities, as a result of neutral hydrogen self-absorption in relatively low-density conditions[18] (Methods). This behaviour is expected when the outflow cools and expands becoming optically thin. The symmetric wings indicate a large covering factor for the ejecta. Expanding nova shells are characterized by similar *BD* values[19] during some stages as well as some of the emission lines detected here. The latter are also found in low excitation nebulosities surrounding outflowing massive stars[20], which show similar Hα equivalent widths in their final and most violent evolutionary phases[21]. This phase is not witnessed after other strong flares displayed early in the outburst (e.g. day 4). However, these events are not followed by such a strong drop in flux as for the case of the major flares preceding the nebular phase (ED Fig. 1).

The time-scale of the optically thick (P-Cyg) to optically thin (extended wings) transition is the diffusion time-scale of an expanding shell with mass $M_{Shell}$, and it is estimated[21] to be $t_{Dif} \approx 23$ days $(1 / R_{15}) (M_{Shell} / M_{\odot})$, where $R_{15}$ is the radius of the spherical envelope in units of $10^{15}$ cm. For $t_{Dif} = 0.002 - 0.1$ days, conservative time scales relevant in the evolution of the BD, we obtain $M_{Shell} \sim 10^{-8} - 10^{-5}$ $M_{\odot}$. This is consistent with the black hole blowing away a significant fraction of the matter stored in its large accretion disc[22] ($\sim 10^{-5}$ $M_{\odot}$). On the other hand, this amount of mass is able to explain the increase in the equivalent hydrogen column density (up to $N_H \sim 1 \times 10^{24}$ cm$^{-2}$) observed during both the 1989 and the 2015 outbursts[22] (Methods).

The active phase of the 2015 outburst of V404 Cyg is much shorter (~15 d) than typically observed in other luminous black holes (months to a year). This is followed by a sharp decay (~3 days), still during the radio-loud phase of the outburst, right after the X-ray peak is reached. This behaviour is consistent with that observed in the 1989 outburst[22]. During these brief outbursts only about 0.1 % ($0.3-1.1 \times 10^{-8}$ $M_{\odot}$) of the material stored in the accretion disc is accreted by the black hole. This corresponds to the gas kept in the innermost $\sim 6 - 9 \times 10^5$ km, which is unaffected by the long-lived outer disc wind. The amount of mass transferred from the donor star to the accretion disc during the preceding 26 years of quiescence is estimated to be $-\Delta M_2 \approx 3 \times 10^{-8}$ $M_{\odot}$ (Methods). This amount is comparable to that accreted by the BH and ejected by the wind. We also detect a prominent double-peaked $H_{\alpha}$ line right after the end of the nebular phase, indicating the presence of a remnant accretion disc once the most active phase of the outburst was finished. Strong $H_{\alpha}$ has in fact been observed throughout the inter-outburst interval, and it is formed at ~80% of the outer disc radius[23] ($0.8 R_{out} = 7 \times 10^6$ km). This (quiescent) period is also characterized by an X-ray luminosity ~2 orders of magnitude brighter than typically observed in quiescent black holes[24], implying ongoing accretion from the remnant accretion disc. Similarly, the much fainter secondary outburst detected in December 2015[25] also indicates the presence of an active disc only 6 months after the major outburst. This time lapse is consistent with the viscous time scale for refilling the inner disc (Methods).

By contrast with the sparse optical data obtained during the 1989 outburst[26], the intense observing campaign presented here allowed us to study in detail the evolution of the wind outflow and to detect the short-lived nebular phase. A nebular phase might have also occurred in the 1989 outburst – where intermittent P-Cyg profiles were detected[26] – but missed because of the scarce monitoring. On the other hand, the relative proximity of V404 Cyg enables both detailed spectroscopic observations at luminosities as low as $10^{-3}$ $L_{EDD}$ and the detection of outflow features as weak as 2% of the continuum level during the brightest phases. In addition, the large accretion disc implied by the 6.5d orbital period – the majority of black holes have orbital periods shorter than ~2d[27] – provides the resource for the formation of outer disc outflows. Besides V404 Cyg, the behaviour of the other two systems with the longest orbital periods might also be influenced by the presence of mass outflows. V4641 Sgr, with $P_{orb}=2.8$d, has shown several brief outbursts characterized by strong radio emission. Extended $H\alpha$ wings ($\pm$ 2,500 km s$^{-1}$) have been reported in a low-luminosity observation, possibly with a weak P-Cyg profile in a Fe II emission line[28]. Likewise, GRS 1915+105, the longest period black hole, has been permanently in outburst for the last 23 years, alternating lower luminosity plateau phases with short luminous episodes lasting only several weeks[4].

It is interesting to note that both V404 Cyg and GRS 1915+105 share some distinctive variability patterns in their X-ray and optical emission regardless of differing luminosities[7]. These include short-term variations with large amplitudes, which, in addition to the neutral wind outflows reported here, appear as a common feature of long period black holes. This variability pattern has been proposed to result from insufficient mass flow reaching the inner parts of large discs, and it might also affect the outburst evolution in addition or alternatively to the presence of the disc outflow presented here. The highly ionized wind of GRS 1915+105 during high accretion rate phases has been suggested to play a role in explaining the variability properties of the source, thereby linking (X-ray) outflows and oscillation patterns[29]. Unfortunately, this system cannot be observed in the optical due to high interstellar extinction. Furthermore, it is not clear whether or not a similar coupling mechanism could be at work at the much lower luminosities associated with both the variability patterns[7] and the neutral wind outflow observed in V404 Cyg.

The sustained disc wind that we have discovered in V404 Cyg could be a new fundamental ingredient in how accretion proceeds in the largest and hence most powerful BH accretion discs. The outflow likely regulates the evolution of the outburst by depleting a sizable fraction of the outer disc, thereby detaching the innermost regions, which are eventually accreted. This suggests analogous behaviour to the cold and massive outflows seen in AGN, which shape their host galaxies at long distances from the central black hole[30].

**Acknowledgments.** Nine of the spectra of 27 June were taken during the visit of His Majesty King Felipe VI of Spain to the 10.4-m Gran Telescopio Canarias (GTC); we appreciate the support this visit provides to astrophysical research in Spain. This work is based on observations made with the GTC telescope, in the Spanish Observatorio del Roque de los Muchachos of the Instituto de Astrofísica de Canarias, during both Time Allocation Committee and Director's Discretionary observing time. We are thankful to the GTC team for the fast response and efficient work throughout the observing campaign. We acknowledge support by the Spanish Ministerio de Economía y competitividad under grants AYA2013-42627 and PSR2015-00397, the Leverhulme Trust Visiting Professorship Grant VP2-2015-046, the International Research Fellowship program of the Japan Society for the Promotion of Science (PE15024), the Bundesministerium für Wirtschaft und Technologie (BMWI/DLR, FKZ 50 OR 1408) and the French Research National Agency's CHAOS project ANR-12-BS05-0009. The use of the MOLLY software developed by T. R. Marsh is gratefully acknowledged.

**Authors Contribution:** T.M-D. performed the GTC data analysis and wrote the paper. J.C. contributed to the GTC data analysis and assisted in writing the paper. D.M.S. performed the GTC data reduction and contributed to the GTC data analysis. R.P.F. provided the radio data and contributed to the scientific discussion. M.A.P. performed X-ray analysis and contributed to the scientific discussion. M. L. provided day-12 GTC spectra and assisted in writing the paper. G.P. contributed to the scientific discussion. P.A.C. contributed to the scientific discussion and assisted in writing the paper. K.P.M. performed radio data analysis. J.R. provided part of the INTEGRAL data.

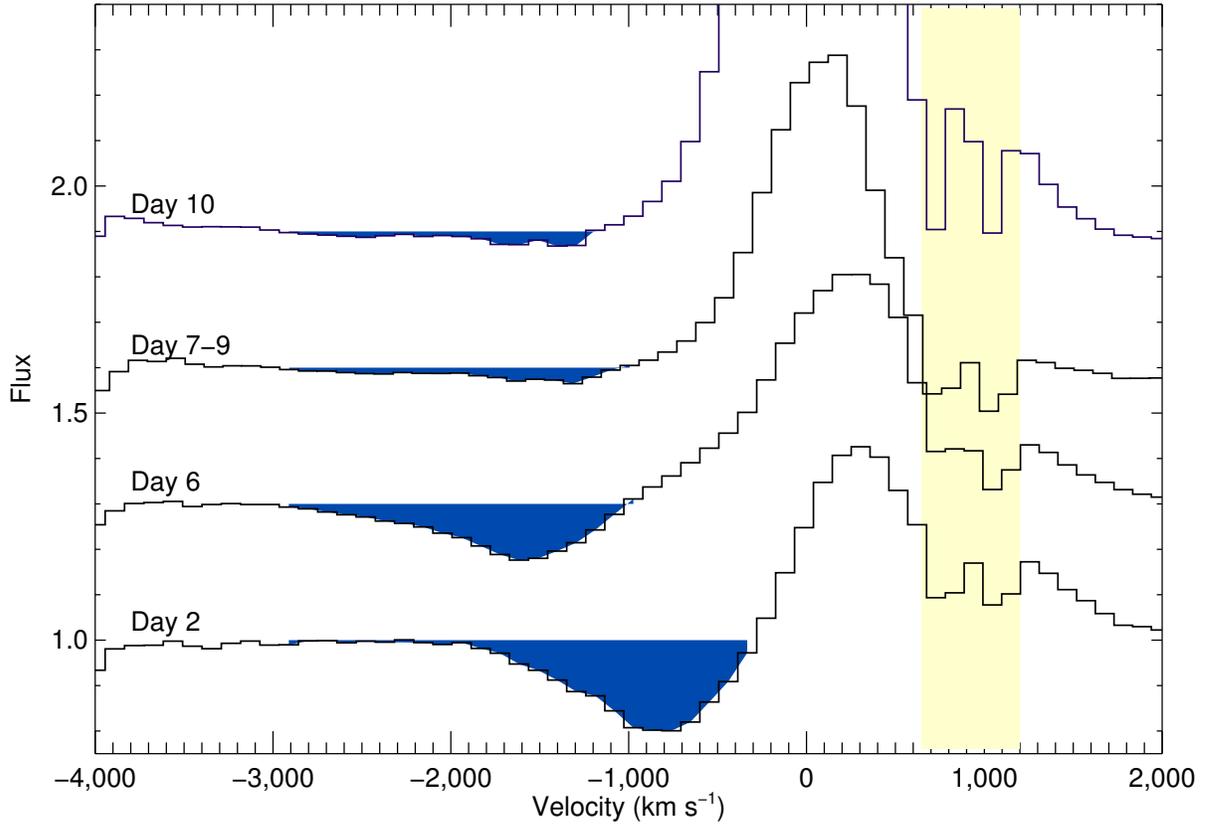

**Fig. 1: P-Cyg profiles observed during days 2, 6, 7-to-9 and 10 in He I 5876 Å.** Normalized spectra are offset by 0, 0.6, 1.2 and 1.8, respectively. Profiles are formed when atomic material approaching the observer at -$V_{out}$ scatters photons with frequency $\nu = \nu_0(1-V_{out}/c)$, while receding ejecta are being illuminated by the central source. Light yellow background indicates regions contaminated by interstellar absorption. We detect approaching material moving up to 3,000 km/s (blue filled absorptions). During days 7 to 9 and 10 the profiles are very shallow, in correspondence with high ionization states (see text). Simultaneously to blue-shifted absorption we detect red-shifted emission detached from the accretion disc component (see also Fig. 2). It reached similar amplitudes than those produced by approaching material, a feature indicating spherical geometry or at least large covering factor[14].

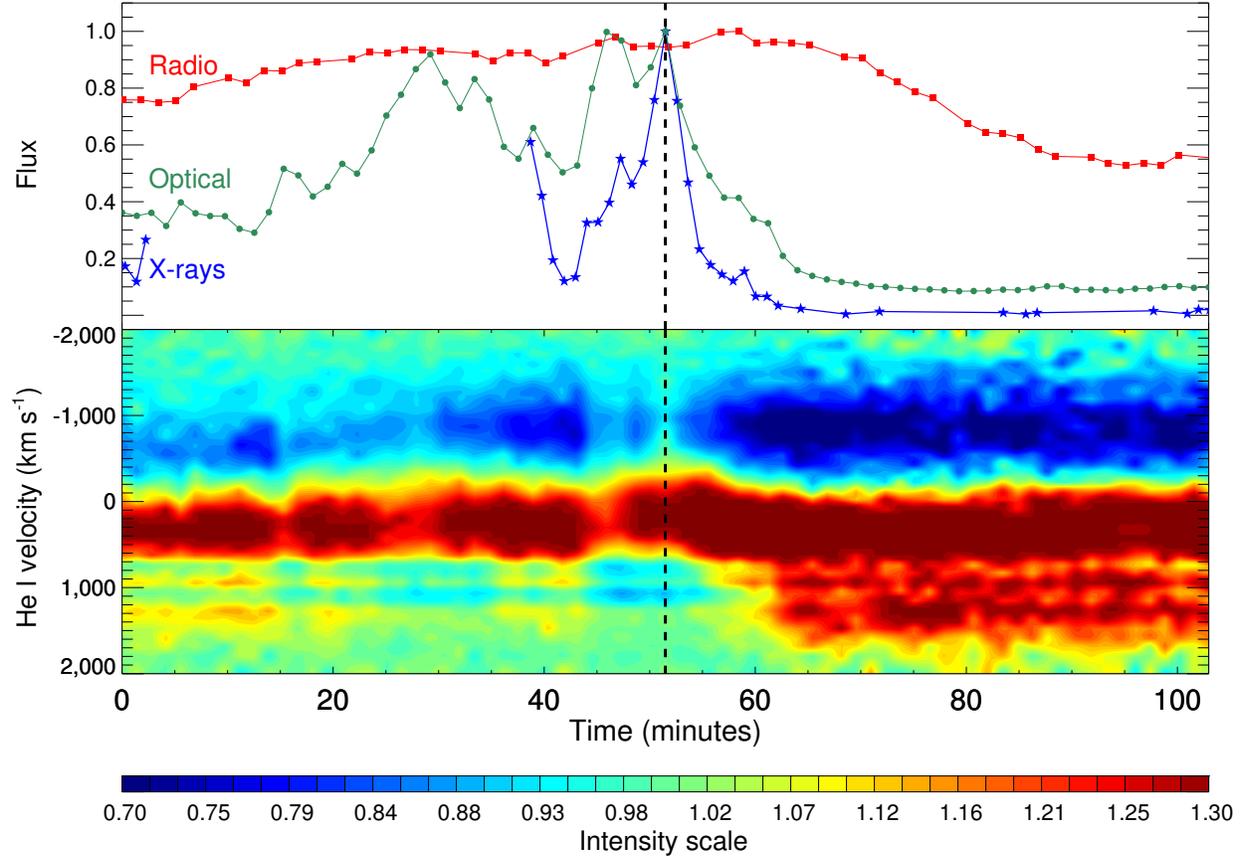

**Fig. 2: Trailed spectrum corresponding Jun 19 data (day 2)**. The trail (bottom panel) covers 103 min with 75 spectra. Time corresponds to minutes from MJD 57192.04. Normalized intensity scale is such that absorptions are represented in blue colours, while emissions are plotted in red. Simultaneous X-ray (*INTEGRAL*; blue stars), optical (green dots) and radio (red squares) normalized light-curves are shown in the top panel. Outflows are detected along the observation, but their properties change in response to flaring. The strongest features become evident right after a sharp X-ray flare is seen (dashed line), as soon as the X-ray flux decreases and $I_{\rm Ratio}$ reaches values as low as 0.5. During the flare (at ~0.08 times the flux peak observed later in the outburst), the P-Cyg profile becomes weaker, as $I_{\rm Ratio}$ increases to values larger than unity (up to ~2).

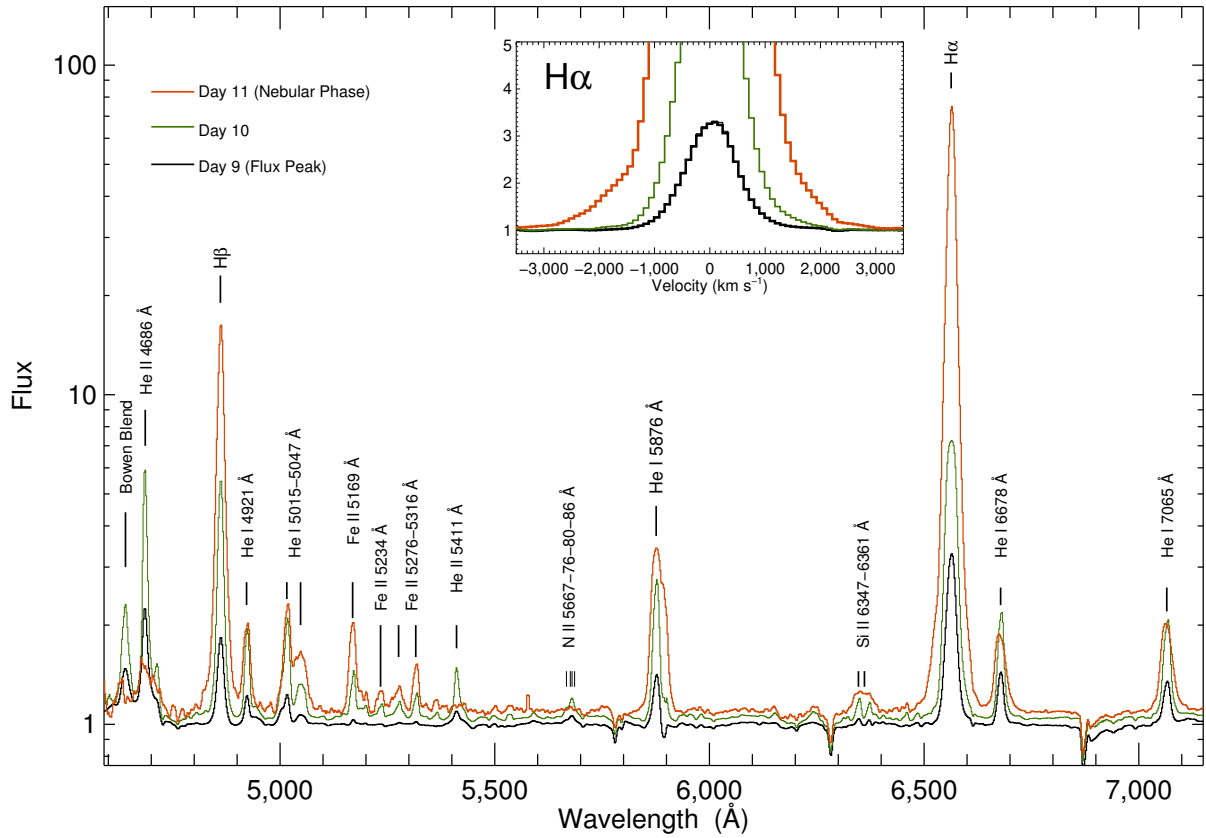

**Fig. 3: Spectral evolution towards the *nebular phase*.** Average, normalized GTC spectra corresponding to days 9 to 11. Log scale has been used due to the intense $H_\alpha$ emission, which reached equivalent width of 2,000 Å. An offset of 0.1 and 0.2 has been added to day 10 and day 11 spectra, respectively. The optical flux drops by 2 orders of magnitude in correspondence to the decay of the X-ray and radio outburst (ED Fig. 1). HeI and, Balmer lines become intense and broad as other transitions become evident (Si II, Fe II). The inset shows a zoom of the $H_\alpha$ region, where broad wings reaching ± 3,000 km s$^{-1}$ become apparent.

# METHODS

**Optical Spectroscopy**

*Observations*

V404 Cygni was observed with the Optical System for Imaging and low-Intermediate-Resolution Integrated Spectroscopy (OSIRIS) located in the Nasmyth-B focus of the 10.4m Gran Telescopio Canarias (GTC), at La Palma (Spain). We used three different optical grisms; R1000B (2.12 Å/pix), R2500R (1.04 Å/pix) and R2500V (0.80 Å/pix), which combined with a 1.0" slit give R = 611, 1485 and 1509, respectively. These cover the spectral ranges: 3,630 – 7,500 Å, 5,575 – 7,685 Å, and 4,500 – 6,000 Å. The complete observational set consists of 545 spectra obtained on 14 different nights within June 17 - July 1 (see ED Table 1). The slit was rotated to PA = 59.15º in order to allow for simultaneous observations of a field star.

*Data analysis*

GTC spectra were bias- and flat field-corrected using IRAF standard routines. The wavelength calibration was performed using Hg-Ar, Ne, and Xe lamps provided by the GTC team. Small velocity drifts (< 20 km s$^{-1}$) due to instrumental flexure were measured from the centroid of the OI 5,577.340 and 6,300.304 Å lines and used to correct the individual spectra. We use routines within MOLLY and IDL to perform further analysis of the spectra. Spectra were initially flux-calibrated relative to the comparison field star included in the slit. The latter was calibrated relative to the spectrophotometric standard *Wolf 1346* using a low-resolution spectrum taken with a 10 arcsec slit during a photometric night at airmass ~1.05. Finally, we applied this flux calibration to the whole database.

*Balmer decrement*

The Balmer decrement (*BD*) was obtained by computing the flux ratio H$_\alpha$ to H$_\beta$ after subtracting the underlying continuum flux and correcting for reddening[31] of $E_{B-V} = 1.3$ . Results are comparable to those obtained from the ratio of the peak intensities of both lines, and results are consistent with Case B recombination (2.5-3) with the exception of days 11 to 15, where values as high as 6 are observed. This epoch is the so-called *nebular phase* (see ED Fig. 1 and 5). We also note that values larger than ~3 are found corresponding to the strongest P-Cyg profiles witnessed on days 1, 2 and 6.

*Ionization State*

Given the dramatic flux changes observed in the X-ray and optical bands, strong changes in the ionization state of the disc are expected as a result of differing irradiation levels. A good, widely used tracker of the variable irradiation of the outer disc is the He II-4686 to H$_\beta$ line flux ratio ($I_{Ratio}$) as it is reddening-independent. Nevertheless, as for the *BD*, we compute this ratio after subtracting the underlying continuum flux and correcting by $E_{B-V} = 1.3$. Not surprisingly, $I_{Ratio}$ is strongly correlated with the optical flux. Studies performed on accreting white dwarfs have shown some distinctive optical properties when $I_{Ratio}$ is larger than unity as a result of a high ionization[32]. In our analysis we find that strong P-Cyg profiles (blue absorption deeper than 5% of the continuum) are always associated with $I_{Ratio} < 1$. The deepest profiles (30% below the continuum) are seen at $I_{Ratio} = 0.5$.

*Hα profile*

A visual inspection of the evolution of the $H_\alpha$ line profile reveals a systematic asymmetry. We have the fitted the $H_\alpha$ profile for each spectrum using a Gaussian model centred at the rest wavelength. We find that the line is redshifted during the first 11 days (top panel in ED Fig. 4), reaching velocities above ~100 km s$^{-1}$ during the first 8 days (i.e. just before the brightest phase of the outburst), in correspondence to the strongest P-Cyg profiles. Similarly, the V/R ratio (defined as the ratio of the blue to red equivalent widths) confirms the line asymmetry observed up to day 11 (bottom panel in ED Fig. 4). Error bars account for variability within every observing window. We interpret this as a result of continuous blue absorption (and extra red emission) present in the line profile, with the more extreme cases leading to P-Cyg profiles. This is also sensitive to the presence of low-velocity outflows, which partially cover more central parts of the blue line profile.

**X-ray Observations**

*Integral Observatory*

V404 Cyg was extensively monitored by the INTEGRAL[33] satellite during the 15 day long outburst starting on MJD 57,190 (June 17, 2015)[34]. The data were acquired during satellite revolutions 1,554 to 1,563. In ED Fig. 1 we present a 25-200 keV light curve obtained with the Imager on Board the INTEGRAL Satellite (IBIS)[35] and the upper layer detector ISGRI in 64s time bins.

The raw data were reduced in a very standard way, using the Off line Scientific Analysis software (OSA) version 10.1, similar to that described in the case of V404 Cyg[11] and Cyg X-1[36] (same region of the sky), respectively. IBIS is a coded mask telescope and the data reduction process is iterative: each active source in a given field projects its own shadow onto the detector, and hence contributes to the overall background of the other sources. Hence to extract scientific products of one specific object one must consider all other active/bright sources within the field.

Our reduction procedure started with the production of sky images and mosaics (obtained from combining data acquired during the same satellite revolution) in the user's pre-defined energy ranges (here we used 20–40 keV, 40–80 keV, 80–150 keV, 150-300 keV) to identify the most active sources over each revolution. In this region of the sky (Cygnus) two persistent sources are very bright and active in the hard X-ray/soft γ-ray domain (~0.2—1 MeV), Cygnus X-1 and Cygnus X-3[36], in addition to V404 Cyg. Occasionally, some other objects may show up (e.g. EXO 2030+375), and are thus included in the reduction process of the revolution concerned.

Light curves were then extracted in time steps of 64~s over the 25–60 and 60-200 keV energy ranges. These energy bands are the same as those selected by the INTEGRAL Science Data Centre quick look analysis facility (http://www.isdc.unige.ch/integral/analysis#QLAsources), providing an independent check and while avoiding any potential saturation effects in the 20–25 keV energy range usually considered.

*Swift Observatory*

Following the initial alert[10], V404 Cyg was monitored with the Swift satellite throughout its outburst until it returned to quiescence. We have analyzed a total of 43 observations acquired with the X-ray Telescope[37] (XRT) taken from 17 days starting on MJD 57,188 (June 15, 2015).

A total of 35 observations were performed in windowed timing (WT) mode, while 8 were taken in photon counting (PC) mode. Observations were processed using the HEASOFT v.6.17 software, in particular the

XRTPIPELINE task. For each observation the 0.5-10 keV spectrum, light curve and image were obtained using XSELECT. We used a circular region of 40 arcsec radius centred at the source position (the inner ~9-11 arcsec were excluded for those observations affected by piled-up). A region of similar size and shape, positioned on an empty sky region was used for the background. We created exposure maps and ancillaryresponse files following the standard Swift analysis threads (http://www.swift.ac.uk/analysis/xrt/), and we acquired the last version of the response matrix files from the HEASARC calibration database (CALDB).

**Radio Observations**

V404 Cyg was observed extensively throughout its 2015 outburst between 13-18 GHz by the AMI-LA radio telescope (Cambridge, UK), operating as part of the University of Oxford's 4 PI SKY (4pisky.org) transients programme. The first data were obtained on MJD 57,188.896 (15 Jun 2015) in robotic response to the Swift trigger 643949. The observations took place within two hours of the Swift trigger, revealing a bright (>100 mJy) and fading radio flare[38]. Subsequently, we continued to monitor the source for up to 10 hours every day for the entire period covered by this report.

Quick-look images of the AMI-LA observations were obtained with the fully-automated calibration and imaging pipeline, AMISURVEY[39]. After the outburst, a more careful calibration and RFI excision of the raw data was done using AMI-REDUCE[40]. The calibrated data was then imported to the CASA package[41,42]. Light curves were extracted in time steps of 1~s in six channels across the 5 GHz bandwidth via vector-averaging of the UV data.

Over the period of 15 days of maximum activity of V404 Cyg, flares with peak flux density of up to 3 Jy at 16 GHz are seen. The rise of the flares is generally optically thick, while the decay is optically thin, consistent with adiabatically expanding blobs of plasma (constituting the jet).

**Fundamental parameters of V404 Cyg**

V404 Cyg is a dynamically confirmed black hole X-ray binary with an orbital period of $P_{orb}$=6.47 days[8]. The black hole mass is in the range $M_{BH} = 8-12$ M$_\odot$, the error budget being dominated by the uncertainty in the orbital inclination[43]. We use a black hole mass of $M_{BH} = 10$ M$_\odot$ and a mass ratio of $q = M_{BH}/M_2$=0.067 in every calculation presented in this paper[43], where $M_2$ corresponds to the donor star mass in M$_\odot$. This results in an orbital separation of $2.2 \times 10^7$ km. The outer accretion disc radius, can be expressed as[44]:

$$R_{out} \cong 1.2 \times 10^{11} M_{BH}^{\frac{1}{3}} P_{orb}^{\frac{2}{3}} \text{ [cm]},$$

where $M_{BH}$ is expressed in M$_\odot$ and $P_{orb}$ in days. We obtain $R_{out}$=9×10$^6$ km. We note that the vast majority of black holes have orbital periods shorter than ~2 days, which results in $R_{out}$=4.1×10$^6$ km using the same values for $M_{BH}$ and $q$.

**P-Cyg Profiles**

We have discovered strong P-Cygni profiles in both hydrogen (Balmer) and helium (He I) emission lines. They result from resonant scattering in an expanding outflow, and are well reproduced by models using a variety of velocity laws and a spherical geometry[14,15].

*Profile fitting*

In order to constrain the velocities associated with the He I-5876 P-Cygni profile, we fitted every

individual spectrum as follows:

(i) We fitted a Gaussian to the central disc emission after masking both the blue P-Cyg absorption and the red high velocity emission bump. This fitted model was subtracted from the data.

(ii) We subsequently fitted a two Gaussian model to the residuals, one in absorption to the P-Cyg and another in emission to the red bump. To avoid degeneracy in the fit, both Gaussians were offset by the same velocity (sign understood) and set to have the same width. The intensities were left as free parameters.

Fits provide a good description of the data for relatively deep profiles, as indicated by our visual inspections. We take the *wind mean velocity* to be the offset velocity, while $V_T$ (the wind terminal velocity) is determined by adding to the latter the half width at 1/10 of the intensity. For days 1, 2 and 6, where strong profiles are present throughout the observations, we were able to track their evolution with time. On day 2 (Fig. 2) outflows are detected during the 2 h observation, but their properties – profile strength and velocity – change in response to flaring. Gaussian fits show a constant terminal velocity ($V_T$) throughout the observation, whilst both the amplitude and mean velocity vary following changes in the optical flux and ionization state. A similar flare-outflow correlation is witnessed on day 6, where we measure $V_T = 3,000$ km s$^{-1}$.

*Possible physical interpretations*

A variety of wind launching mechanisms have been proposed in the literature. Here, we briefly discuss the three more widely used:

(i) Radiation pressure: wind results from Thompson scattering when radiation field approaches $L_{EDD}$. Given the low luminosity at which the most conspicuous P-Cyg profiles are detected (0.1−1%$L_{EDD}$), and the similar $V_T$ values observed during the outburst, radiation pressure is probably not responsible for the observed phenomenology. However, the system might have reached $L_{EDD}$ during the brightest phases of the outburst preceding the so-called *nebular phase,* and this mechanism could have contributed to the observed optically thin shell. On the other hand, we note that the velocity observed during this phase is similar to that measured in the P-Cyg profiles, suggesting a common origin.

(ii) Line-driven winds: expected to be inefficient in LMXBs since X-ray emission from the disc would over-ionise the wind[45].

(iii) Thermal wind scenario[16]. In this scenario, atoms reach a thermal velocity larger than the escape velocity and a wind is formed. Therefore, the launching radius $R_l$ is approximately that where an associated Keplerian velocity equals $V_T$. For $V_T = 3,000 - 1,500$ km s$^{-1}$ we obtain $R_l = 1.5 - 6 \times 10^5$ km, respectively.

Using a standard accretion disc model[46] we can estimate the surface temperature at a given radius $R$ by:

$$T_D(R) = \left[\frac{3GM_{BH}M_{acc}}{8\pi R^3 \sigma}(1 - \sqrt{\frac{R_0}{R}})\right]^{1/4},$$

where $G$ is the gravitational constant, $\sigma$ the Stefan–Boltzmann constant, and $R_0$ the disc inner radius

$$R_0 = \frac{6GM_{BH}}{c^2}$$

The accretion rate is obtained using:

$$\dot{M} \approx 2.0 \times 10^{39} \frac{L_{out}}{\eta c^2},$$

where $L_{out}$ is the Eddington scaled luminosity at the time of the outflow, $c$ is the speed of light and $\eta=0.1$ the accretion efficiency. Using $L_{out} = 0.001 - 0.01 L_{EDD}$ we obtain $T_D$ in the range $5,000-20,000$ K, which is consistent with having both neutral hydrogen and helium.

*Mass outflow rate*

A crude estimate of the mass loss can be obtained by direct comparison of the strongest profiles (day 2) with the classical atlas of theoretical P-Cyg profiles[15]. Following studies of cataclysmic variables[47] we used:

$$\dot{M}_{out} \approx 1.1 \times 10^{-18} \frac{\tau R V_T^2}{f_i A \lambda_0 g} C \; [M_\odot \, yr^{-1}],$$

where $\tau$ is the average opacity, $\lambda_0$ the rest wavelength, $f_i$ the ionization fraction, $A$ the helium abundance, $R$ the radius of the emitting region in solar radii and $g$ the oscillator strength. $C$ is an integral depending on the shape of the P-Cyg profile, which takes typical values in the range 0.2 to 0.5. By visual inspection we set $\tau \approx 2$ by comparing our profiles with those of the atlas.

Using $V_T = 2,000$ km s$^{-1}$, $g=0.61$, $C=0.2$ and $A=0.08$ we obtain $\dot{M}_{out} > 1 \times 10^{-14} M_\odot \, yr^{-1}$. Note that this is a lower limit because (i) we have assumed $f_i=0.5$, which might be much lower depending on e.g. the exact value of the wind launching radius (although the low *He II-4686 / Hβ* intensity ratio $I_{Ratio} =0.5$ advocates for a significant fraction of neutral helium), and (ii) we used R=6×10$^5$ km for the proposed launching radius, which could be significantly larger if e.g. the shell remains optically thick ~300 s (0.003 days) after the outflow is launched. Indeed, if the shell is optically thick for at least 0.01 days (see below), we obtain $\dot{M}_{out} > 1 \times 10^{-13} M_\odot \, yr^{-1}$ ($C= 0.5$).

On the other hand, the contemporaneous Eddington-scaled accretion rate for $L_{out} = 0.001 L_{EDD}$ would be $\dot{M}_{acc} \approx 10^{-10} M_\odot \, yr^{-1}$ and then $\frac{\dot{M}_{out}}{\dot{M}_{acc}} > 10^{-3}$.

*Nebular Phase*

From days 9 to 11 we observe major changes in the spectrum, as the X-ray, radio and optical fluxes drop by 2—3 orders of magnitude from the outburst peak: (i) A weak P-Cyg profile is still present on days 9 and 10, which means that the ejecta is optically thick; (ii) higher excitation emission lines become strong on day 10 and Balmer line equivalent widths start to increase; (iii) on day 11, emission lines become unprecedentedly broad and intense, showing zero intensity breadths of ~6,000 km/s and equivalent widths of ~ 2,000 Å (H$_\alpha$; Fig. 3, ED **Fig**. 1). This results from material expanding at the outflow velocity (± 3,000 km s$^{-1}$) and becoming optically thin, as typically observed in expanding nova shells. We note that H$_\alpha$ saturated the detector on day 10, so its intensity has to be taken as a lower limit. Similar spectra to that presented in Fig. 3 (day 11) were observed during days 12, 13 and 14, although line intensities progressively decay. Day 15 data show features typical of quiescent black hole transients, including a double peaked H$_\alpha$ emission line.

*Diffusion time scale and ejected mass*

The time-scale of this transition is the diffusion time-scale of an expanding shell with mass $M_{Shell,}$ and it is estimated[21,48] to be $t_{Dif} \approx 23$ days $(1 / R_{15})(M_{Shell} / M_\odot)$, where $R_{15}$ is the radius of the spherical envelope in units of $10^{15}$ cm. The transition from optically thick to optically thin ejecta occurs between day 10 and 11, as we observe a P-Cyg profile and $BD\sim2.5$ on day 10 and broad wings and $BD\sim5$ on day 11 (ED Fig. 5). This means that the outflow becomes optically thin in $t_{Dif}$ <1 days if ejection of matter continues up to day 10. On the other hand, on day 11 we have two separate groups of observations with $BD$ increasing across ~0.01 d time scales, which suggests that this time-scale could be relevant in the expansion. Extrapolating from this variation, we predict a maximum $t_{Dif}\sim0.1$ days. On the other hand, we do not observe significant changes from spectrum to spectrum ($t_{Dif}$>0.002). Assuming a $3\times10^5$ km launching radius and material travelling at 3,000 km s$^{-1}$ we obtain $R_{15}= 9\times10^{-5} - 3\times10^{-3}$. This yields $M_{Shell} = 10^{-8} - 10^{-5}$ M$_\odot$ for $t_{Dif}\sim0.002-0.1$, respectively. This order-of-magnitude calculation is consistent with the amount of matter expected to be stored in a large accretion disc such as that of V404 Cyg ($\sim10^{-5}$ M$_\odot$)[22]. Similarly, it also explains nicely why the optically thin nebulae is detected right after the end of the outburst (<1 day), a much shorter time-scale than typically observed in supernovae (tens to hundreds of days) where >1—10 M$_\odot$ are expelled[21]. The above estimates are quite approximate (at least a factor of ~2) and they assume a spherical geometry. Nevertheless, our results are consistent with a significant fraction of the disc being ejected during the outburst. On the other hand, an increase in the equivalent hydrogen column density ($N_H$ up to a few times $\sim10^{23}$) has been reported for both the 1989[22] and 2015[49] outbursts. Assuming a spherical geometry for the wind and constant density across the outflow we predict $N_H\sim10^{21} - 10^{24}$ cm$^{-2}$ if $M_{Shell}$ ($10^{-8} - 10^{-5}$ M$_\odot$) was expelled.

*Mass Transferred by the donor during the inter-outburst period*

Given the long orbital period of V404 Cyg, the mass transfer rate from the donor can be estimated using the following expression[44]:

$$-\dot{M}_2 \approx 4.0 \times 10^{-10} P_d^{0.93} M_2^{1.47} \text{ [M}_\odot \text{ yr}^{-1}]  ,$$

where $P_d$ is the orbital period in days. This results in $-\dot{M}_2 \approx 1.3 \times 10^{-9}$ M$_\odot$ yr$^{-1}$, which translates into $-\Delta M_2 \approx 3\times10^{-8}$ M$_\odot$ across the 26 yr long inter-outburst period.

*Accreted mass estimate*

The total mass accreted during the 2015 outburst based on the observed X-ray luminosity can be estimated using the following expression:

$$\Delta M_X \approx \frac{\int L_X}{\eta c^2}  ,$$

where $\int L_X$ is the integrated X-ray luminosity throughout the outburst. This was estimated by converting the observed Integral count rate to flux in the 10 keV – 1 MeV band (https://heasarc.gsfc.nasa.gov/cgi-bin/Tools/w3pimms/w3pimms.pl). We assumed[11] a Power-law spectral model with a photon index in the range $\Gamma=1-2$ and $N_H\sim0.7\times10^{22}$, even though the results does not depend on the later value. Using $\eta=0.1$ we obtain $\Delta M_X \approx 0.6-2.3 \times10^{25}$ g (i.e. $0.3-1.1 \times10^{-8}$ M$_\odot$). The mass implied by the soft X-ray luminosity (0.5—10 keV), and thereby sensitive to (variable) absorption effects, is estimated from both Swift and Integral (extrapolation) to be only a few times $10^{23}$ g. Our results are compatible with $\Delta M_X \sim 3 \times10^{25}$ g inferred during the 1989 outburst[22], showing that only $\sim0.5-1\times10^{-3}$ of the total mass stored in the disc ($R_{out}=9\times10^6$ km) was accreted. The disc mass[22, 50] varies as $\sim R_{out}^3$, which in turn implies that the accreted mass corresponds to that within $R_{Acc} \approx 6 - 9\times10^5$ km

Renewed activity of V404 Cyg[12, 51], at much fainter level, was detected during Dec 23, 2015 – Jan 5, 2016, i.e. only ~6 months after the main 2015 outburst. This time scale is consistent with the viscous time for refilling $R_{Acc}$ as compared with accreting white dwarfs. This can be expressed as:

$$t_V \approx N_{WD}\, 2.87\times10^7\, R_{10}^{0.61} M_{BH}^{0.46}\ ,$$

where $R_{10}$ is the radius in units of $10^{10}$ cm and $N_{WD}$ was (roughly) calibrated to be ~0.05 using observations of accreting white dwarfs[52]. Using $R_{Acc} = 6 - 9\times10^5$ km, we obtain $t_V$ ~140−180 d, which is compatible with the time lapse between the two outburst.

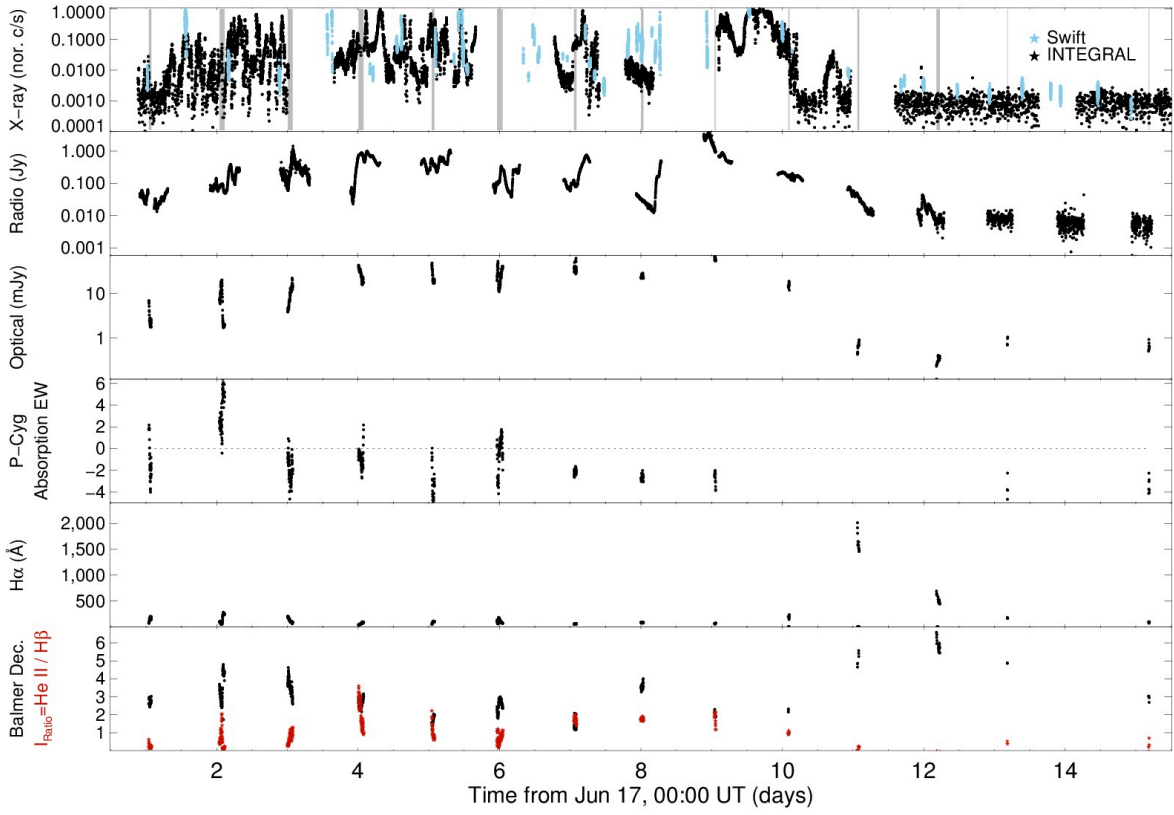

**Extended Data Fig. 1:** Evolution of the main parameters during the the outburst. Zero time is set to Jun 17, 00:00 UT. From top to bottom we show hard (25−200 keV) and soft (0.5 – 10 keV) normalised X-ray count rates, radio flux (∼16 GHz), optical continuum flux, He I 5876 Å equivalent width (positive for absorption) in the range -3,000 – 0 km s$^{-1}$ and H$_\alpha$ equivalent width. Balmer decrement (black) and $I_{\rm Ratio}$ (red) are shown in the bottom panel. In the top panel, X-rays have been normalized to their respective peak at ∼L$_{\rm EDD}$ and the time intervals corresponding to the GTC observations have been greyed out for clarity.

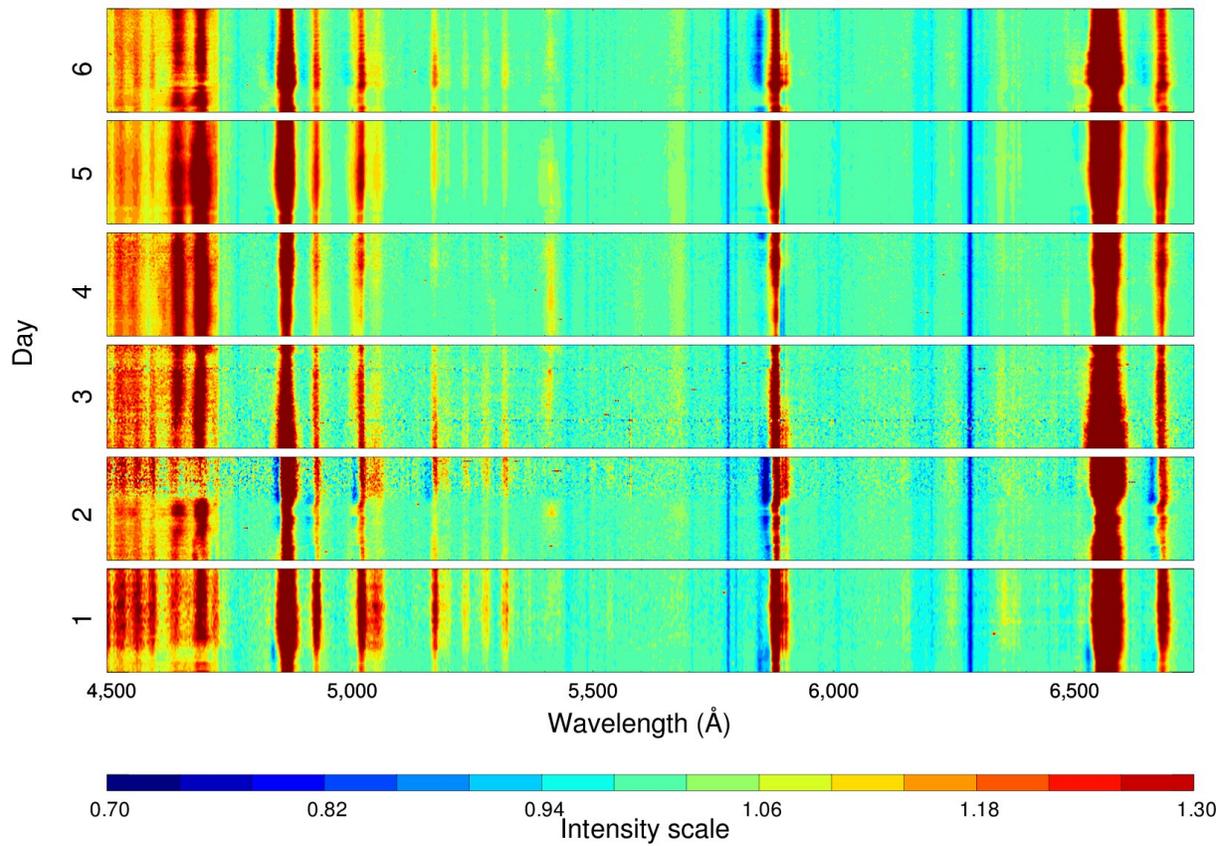

**Extended Data Fig. 2: Trail spectrum showing GTC spectra taken during days 1 to 6.** P-Cyg profiles are apparent in seven transitions of neutral hydrogen (H$\alpha$ and H$_\beta$) and helium. The strongest are observed in days 1,2 and 6, being more prominent in the He I 5876 Å transition. Similar profiles are seen in another 5 transitions at shorter wavelengths (not shown).

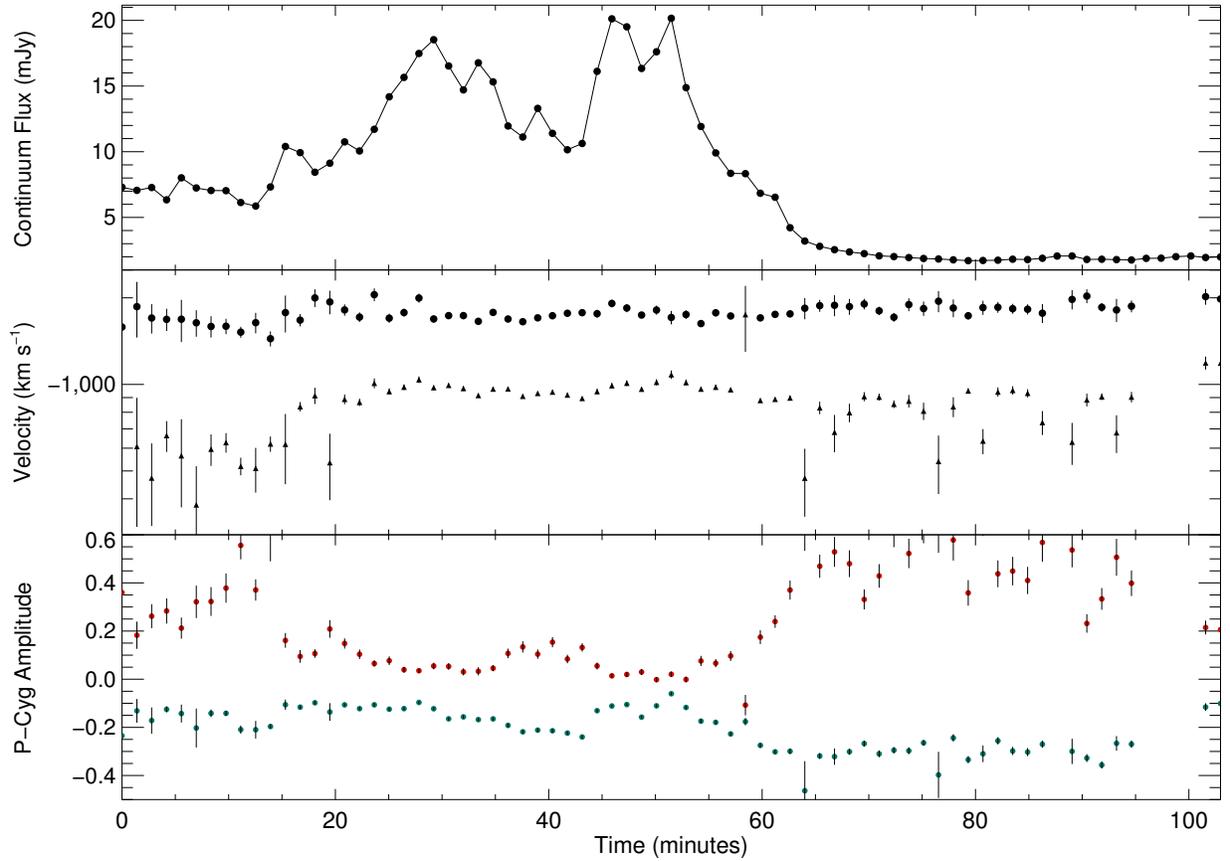

**Extended Data Fig. 3: Gaussian Fits to the P-Cyg profiles.** From top to bottom we show the optical flux, terminal (dots) and mean (triangles) velocities, and emission (red dots) and absorption amplitudes (blue dots) of the P-Cyg profiles. The two bottom panels result from a Gaussian fitting after subtracting the disc component from the emission (Methods). Terminal velocities in the range $V_T=1,500 - 2,000$ km s$^{-1}$ are witnessed (see Fig 2.). This method yields $V_T=3,000$ km s$^{-1}$ for day 6 (see Fig. 1). The amplitude of the profile is correlated with the optical flux. The similar (if not higher) amplitude of the emission component implies that the wind has a large covering factor. Error bars indicate the standard error of the mean.

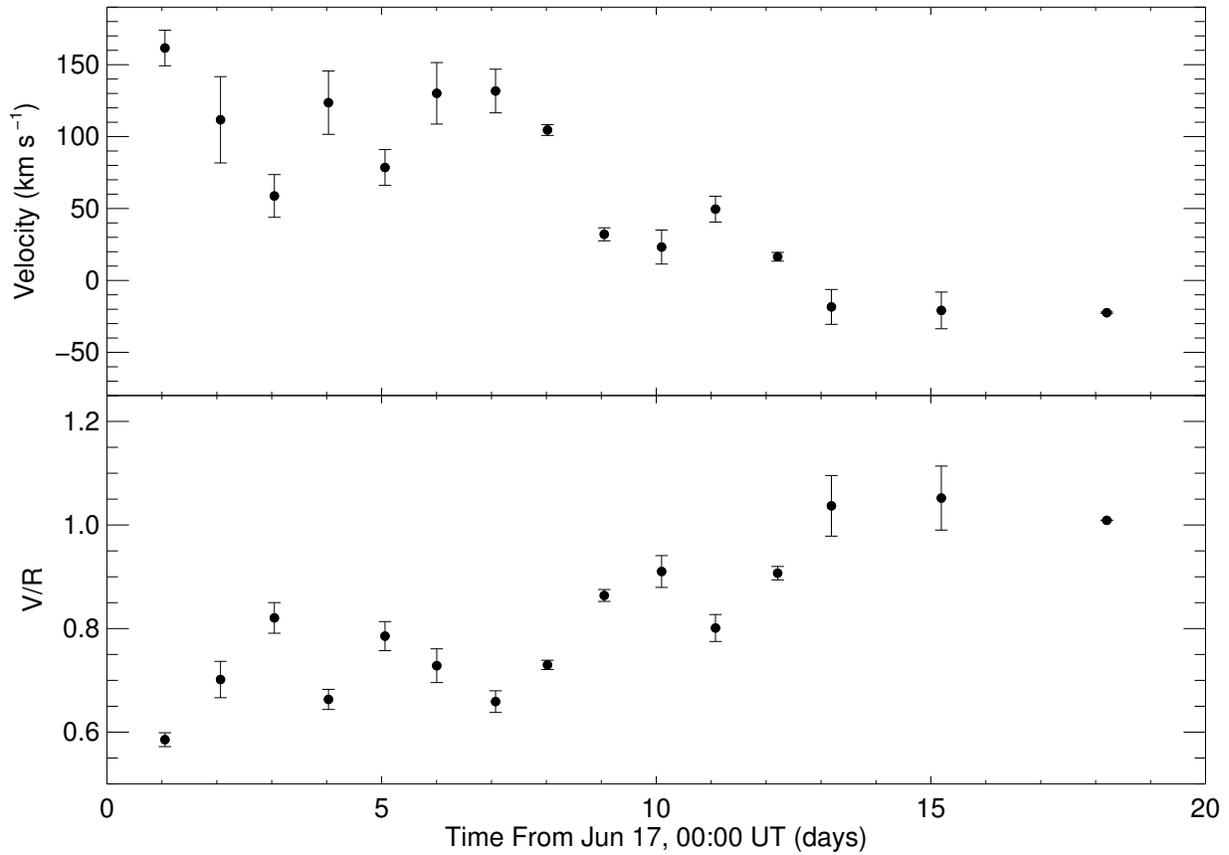

**Extended Data Fig. 4:** top panel: evolution of the centroid of the Hα line. Positives values are due to line asymmetries by blue absorptions and red emission. Bottom panel: V/R parameter (symmetric profile if 1) showing the same phenomena. This strongly suggests the presence continuous outflows from the outer disc along the whole outburst. Error bars indicate the standard deviation of measurements within each observing window.

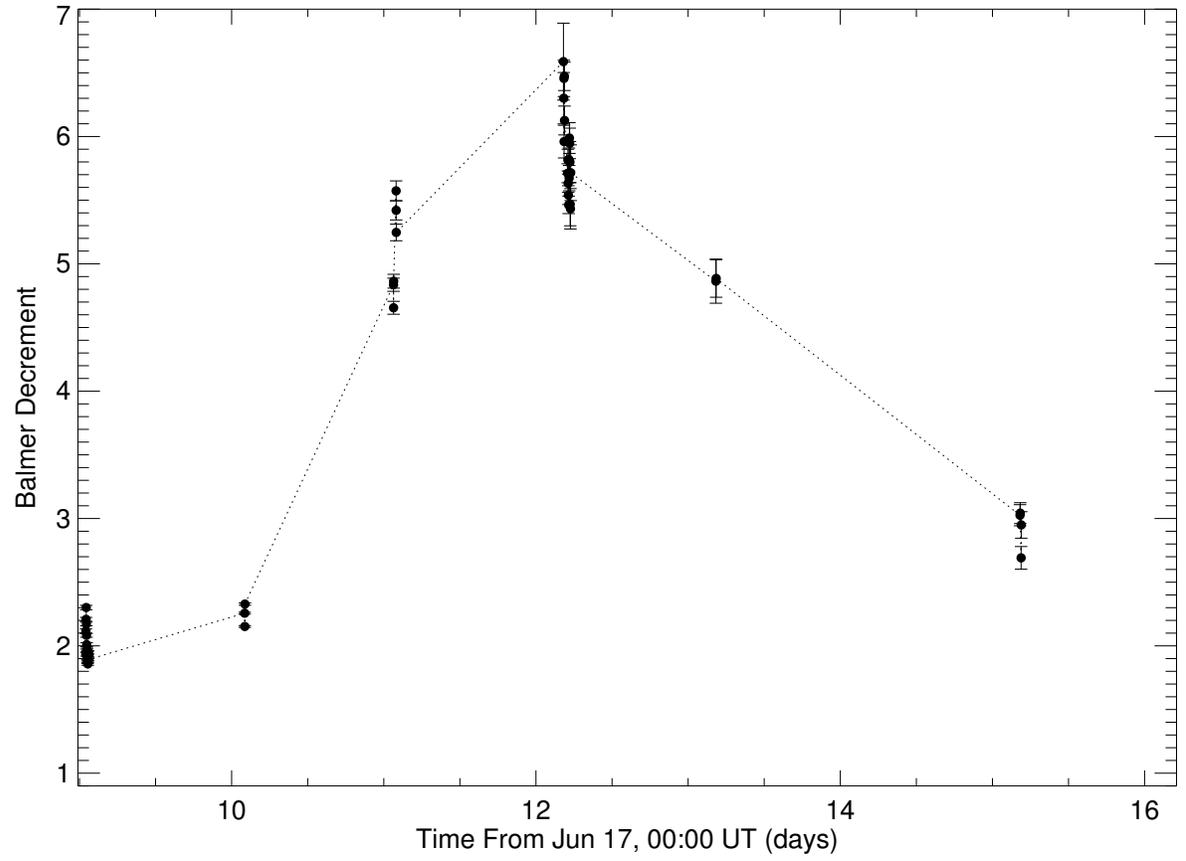

**Extended Data Fig. 5: Balmer decrement evolution through the nebular phase.** From day 10, the Balmer decrement is observed to sharply increase, reaching ~5 on day 11 and ~6 on day 12 (see Methods). Error bars indicate the standard error of the mean.

| Night | Grism | $T_{EXP}$ [s] | $TR$ [s] | $N_{SPEC}$ |
|---|---|---|---|---|
| 17-06-2015 | R1000B | 60 | 84 | 36 |
| 18-06-2015 | R1000B | 60 | 84 | 75 |
| 19-06-2015 | R1000B | 60 | 84 | 75 |
| 20-06-2015 | R1000B | 60 | 84 | 75 |
| 21-06-2015 | R1000B | 60 | 84 | 40 |
| 22-06-2015 | R1000B | 60 | 84 | 85 |
| 23-06-2015 | R1000B | 60 | 84 | 36 |
| 24-06-2015 | R1000B | 60 | 84 | 36 |
| 25-06-2015 | R1000B | 60 | 84 | 17 |
| 26-06-2015 | R1000B | 40 | 64 | 3 |
|  | R2500R | 70 | 94 | 6 |
|  | R2500V | 70 | 94 | 6 |
| 27-06-2015 | R1000B | 20 | 44 | 3 |
|  | R2500V | 70 | 94 | 3 |
|  | R2500R | 35 | 59 | 3 |
|  | R1000B | 20 | 44 | 3 |
|  | R2500V | 70 | 94 | 3 |
|  | R2500R | 35 | 59 | 3 |
| 28-06-2015 | R1000B | 20 | - | 1 |
|  | R1000B | 60 | 84 | 5 |
|  | R2500V | 360 | 384 | 3 |
|  | R2500R | 120 | 144 | 3 |
|  | R1000B | 120 | 144 | 13 |
| 29-06-2015 | R1000B | 120 | 144 | 2 |
|  | R2500R | 120 | 144 | 2 |
| 01-07-2015 | R1000B | 120 | 144 | 2 |
|  | R2500R | 120 | 144 | 2 |
|  | R1000B | 60 | 84 | 2 |
|  | R2500R | 60 | 84 | 2 |

**Extended Data Table 1. Log of the GTC observations.** $T_{EXP}$ correspond to the exposure time per spectrum in seconds, TR is the actual time resolution and $N_{SPEC}$ the number of spectra taken on a given day and with a given configuration.